\documentclass[aip,apl,reprint,twocolumn]{revtex4-1}
\usepackage{graphicx}
\usepackage{dcolumn}
\usepackage{bm}
\usepackage{amssymb,amsmath}
\usepackage{gensymb}
\usepackage{balance}
\usepackage{textcomp}
\usepackage{times,mathptmx}
\usepackage{graphicx}
\usepackage{lastpage}
\bibstyle{apsrev}
\usepackage{hyperref}
\usepackage{color}
\usepackage[normalem]{ulem}

\begin{document}
\preprint{PREPRINT}

\title[Study of In-plane and Interlayer Interactions During Aluminum Fluoride Intercalation in Graphite: Implications for the Development of Rechargeable Batteries]
{Study of In-plane and Interlayer Interactions During Aluminum Fluoride Intercalation in Graphite: Implications for the Development of Rechargeable Batteries}


\author{Sindy J. Rodríguez}
\email{sindy.rodriguez@santafe-conicet.gov.ar}
\author{Adriana E. Candia} 
\affiliation{Instituto de F\'isica del Litoral, Consejo Nacional de Investigaciones Científicas y Técnicas y Universidad Nacional del Litoral (IFIS-Litoral, CONICET-UNL) Santa Fe, 3000, Argentina}
\author{Igor Stankovi\'{c}}
\email{igor.stankovic@ipb.ac.rs}
\affiliation{Scientific Computing Laboratory, Center for the Study of Complex Systems, Institute of Physics Belgrade, University of Belgrade, Zemun, Serbia}
\author{Mario C.G. Passeggi (Jr.)}
\affiliation{Instituto de F\'isica del Litoral, Consejo Nacional de Investigaciones Científicas y Técnicas y Universidad Nacional del Litoral (IFIS-Litoral, CONICET-UNL) Santa Fe, 3000, Argentina}
\affiliation
{Facultad de Ingenier\'ia Qu\'imica, Universidad Nacional del Litoral, Santa Fe, 3000, Argentina}
\author{Gustavo D. Ruano}
\affiliation{Centro At\'omico Bariloche, Comisi\'on Nacional de Energ\'ia At\'omica (CNEA), San Carlos de Bariloche, 8400, Argentina}
\affiliation{Consejo Nacional de Investigaciones Cient\'ificas y T\'ecnicas (CONICET), San Carlos de Bariloche, 8400, Argentina}

\date{\today}


\begin{abstract}
The electrolyte intercalation mechanism facilitates the insertion/extraction of charge into the electrode material in rechargeable batteries. Aluminum fluoride (AlF$_{3}$) has been used as an electrolyte in rechargeable aluminum batteries with graphite electrodes, demonstrating improved reversibility of battery charging and discharging processes; however, the intercalation mechanism of this neutral molecule in graphite is so far unknown. In this work, we combine scanning tunneling microscopy (STM) in ultra-high vacuum conditions, calculations based on density functional theory, and large-scale molecular dynamics simulations to reveal the mechanism of AlF$_{3}$ intercalation in highly oriented pyrolytic graphite (HOPG). We report the formation of AlF$_{3}$ molecules clusters between graphite layers, their self-assembly by graphene buckling-mediated interactions, and explain the origin and distribution of superficial {\it blisters} in the material. Our findings have implications for understanding the relationship between the mobility and clustering of molecules and the expansion of the anode material. This, in turn, paves the way for future enhancements in the performance of energy storage systems.
\end{abstract}


\maketitle

\section{Introduction}

Renewable and sustainable energy storage technologies are nowadays a strategy to mitigate climate change, environmental pollution, and fossil fuel scarcity. \cite{Gu2019,Steele2001} Electrochemical energy storage, particularly rechargeable batteries, is considered one of the solutions to supply or back up clean electricity in portable devices. While lithium-based rechargeable batteries are developed for various applications and successfully commercialized, much research has focused on exploring alternative materials that are abundant in nature and less reactive, which reduces self-ignition risk. \cite{Zhu2019,Yang2011} In this sense, research on alternatives to lithium (Li) in rechargeable batteries is dominated mainly by systems with sodium (Na),~\cite{Cui2017,Palomares2012} magnesium (Mg) \cite{Saha2014,Aurbach2000} or aluminum (Al) \cite{Lin2015,Leisengan,Yoon2022}. The anode material is equally important to charge carrier ions. Graphite shows potential as an anode material for rechargeable metal-ion batteries because of its high abundance and low cost~\cite{Li2019,Li2020}. The systems with superior rate performance and cycling stability are obtained through a unique potassium-solvent co-intercalation mechanism in natural graphite~\cite{Li2020}. Recently, graphite intercalation compounds (GICs) involving aluminum ions (Al-GIC) have emerged as a promising type of rechargeable batteries due to their high gravimetric density, lower reactivity, and easy handling.~\cite{Ambroz2017,Agiorgousis} Aluminum ion rechargeable batteries (AIBs) have the redox property of involving three electrons during electrochemical processes resulting in a higher volumetric energy density than Li batteries, which is attracting increasing attention from researchers. \cite{Ying2014,Jayaprakash2011,Lin2015,LiuS2012,Rani2013} While the most investigated compounds in this class are AlCl$_3$~\cite{CohnG2015,JIAO2016276,Mandai,Huali2015,Huali2016}, recently, aluminum fluoride AlF$_3$ is proposed as a potential candidate for electrolytes in batteries composed of graphite cathodes and aluminum anodes.~\cite{Wang2017,ZhangM2019} However, one of the main challenges lies in understanding the mechanisms governing the electrochemical performance of graphene anodes for rechargeable batteries ~\cite{Li2019,Li2020}. Infrared spectroscopy and X-ray diffraction~\cite{Li2020} measurements have demonstrated the intercalation and co-intercalation mechanism of large molecular complexes. Yet a complete outline of the interplay of graphene interactions with molecules and between the molecules remains unclear. The reason for this is the inaccessibility of the system for in-situ measurements in well-defined conditions. 

In this paper, we focus on graphite as an anode material and AlF$_3$ molecule as an electrolyte applying a two-fold multiscale approach based on the systematic nondestructive experimental investigation of the intercalation process using scanning tunneling microscopy (STM) under ultra-high vacuum (UHV) conditions and large scale molecular dynamics simulations of multilayer graphite surface, complemented with density functional theory (DFT) calculations. While a complete outline of intricate interactions of various ions and molecular complexes remains elusive, the comprehensive findings concerning AlF$_3$ could provide valuable insights that can be extended to understand the intercalation process for a broader class of materials. On its own, AlF$_3$ has been proposed as an electrolyte in Al batteries with graphite cathodes, since cyclic voltammetry (CV) showed that it protects the electrode during the cycling process, as well as, improves the reversibility, durability, and charge transfer of the device.~\cite{WangQ2019,Wang2017,ZhangM2019} The computational results suggest that using the AlF$_{4}^{-}$ anion instead of AlCl$_{4}^{-}$ could increase the specific capacity and operating voltage of an eventual rechargeable battery. \cite{WangQ2019} A notable experimental study of AlF$_3$ by Wang \textit{et al.} \cite{Wang2017} documented a compelling phenomenon: the introduction of a small quantity of AlF$_3$ into the liquid electrolyte resulted in improved battery reversibility. The potential utilization of AlF$_3$ molecules as guest species within the graphite matrix not only provides a platform for understanding solvent attributes of AlF$_3$ but represents a route for comprehending its incorporation dynamics. Recently, we found that AlF$_3$ molecules are incorporated through the steps-edges present on the HOPG surface, locally separating the carbon layers from the substrate without forming chemical bonds and altering its local density of states (LDOS). ~\cite{Candia2021} Our theoretical study on the bulk, ~\cite{Sindy2021} showed that the AlF$_3$ molecule is energetically unstable on HOPG surfaces and instead intercalates in stable non-planar configuration due to attractive van der Waals forces between the HOPG layers. Despite the progress in the study of AlF$_3$, the mechanism of intercalation of this molecule as a function of the concentration of intercalating molecules should be understood.
\begin{figure*}[th!]
\centering
\includegraphics[width=2\columnwidth]{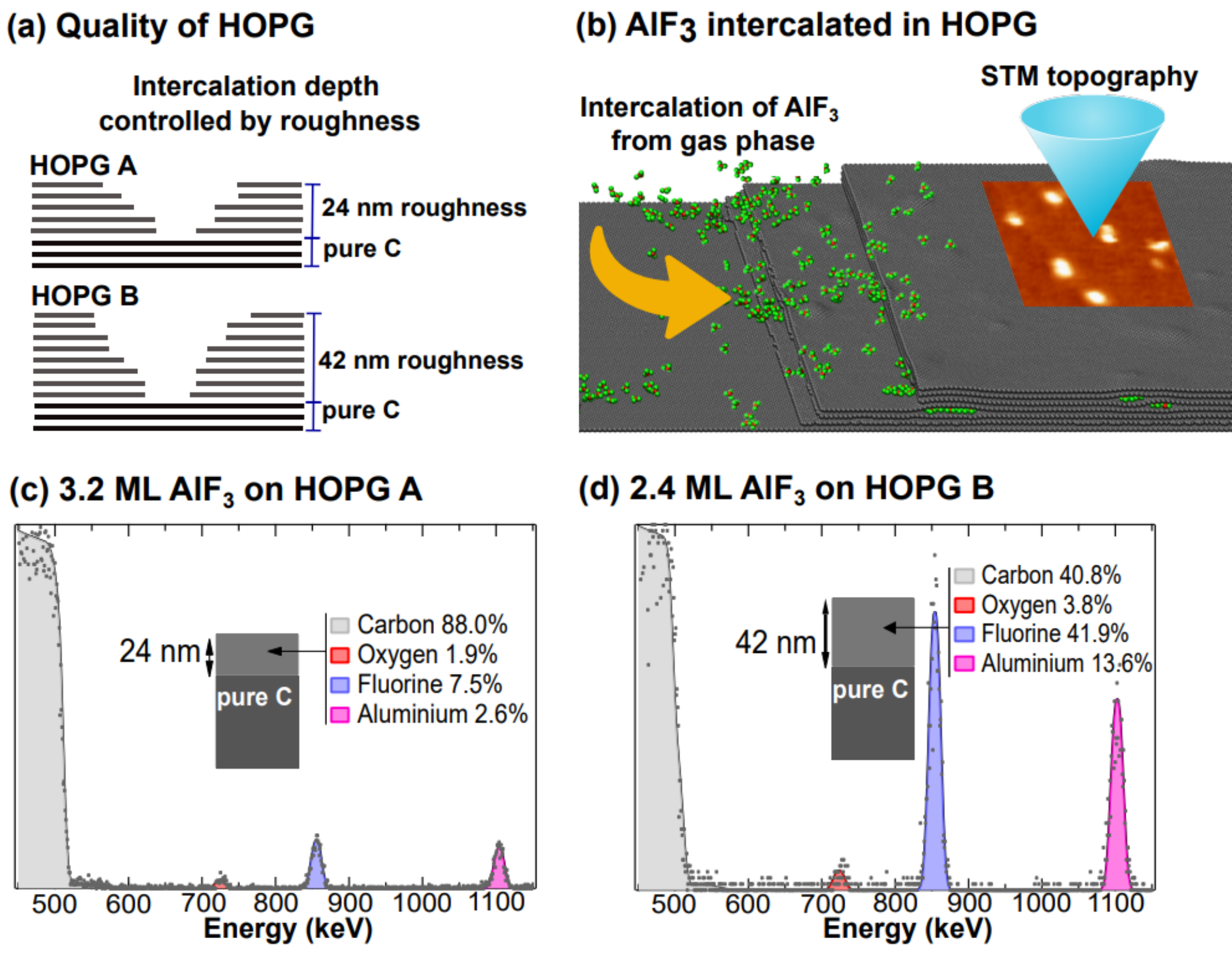}
\caption{(a) Schematic side view of the different HOPG roughness quality used in the RBS experiments to determine the dependence of the vertical penetration depth of AlF$_{3}$ on the topology. (b) STM-UHV preparation and characterization of AlF$_{3}$ intercalated in the HOPG. (c) and (d) RBS experimental spectra (represented by black squares) and fit (full line). In their insets: a bullet of the model obtained from each fit indicates the atomic percentage and depth of the intercalations. Schemes (a), (c), and (d) adapted from Candia et al.\cite{Candia2021}.}
  \label{scm:1}
\end{figure*}

\begin{figure*}[ht!]
\centering
  \includegraphics[width=1.8\columnwidth]{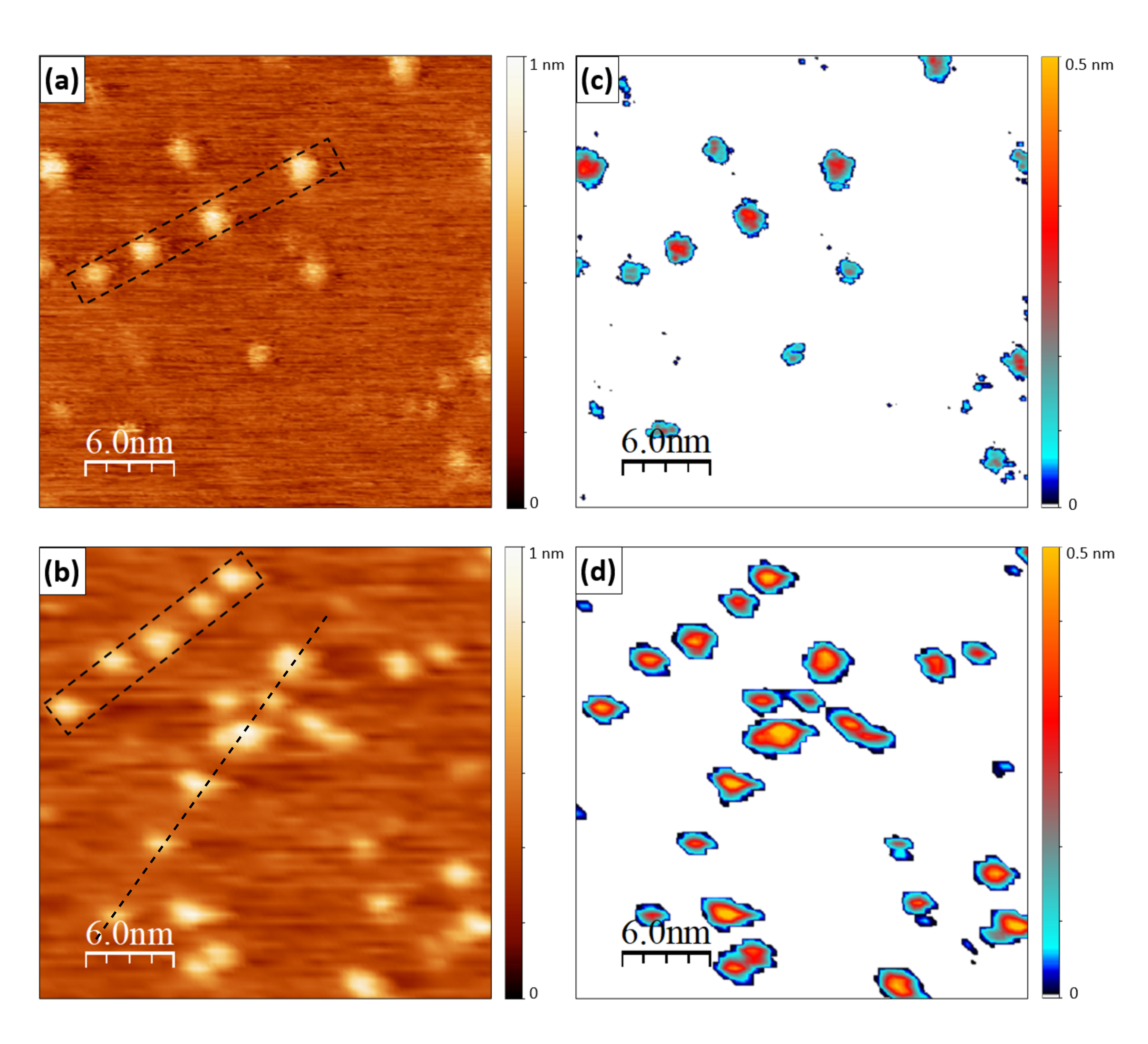}
  \caption{Topography STM images (30~nm $\times$ 30~nm) of HOPG  after the deposition of (a) 1.2 and (b) 3~ML of AlF$_3$ at 300~K. Both images were acquired with a sample bias voltage of V$_{S}$=500, 375~mV and tunnel currents I$_{\rm T}$=0.001, 0.002~nA, respectively. (c) and (d) Height threshold obtained by applying the “flooding” procedure to images (a) and (b).}
  \label{fgr:1_STM}
\end{figure*}

The results presented in this study support the concept of mixed staging for the intercalation of AlF$_3$ in graphite. We explore the topographical and electronic properties of AlF$_3$ molecules intercalating between HOPG layers. The STM images are acquired at room temperature (RT) for different exposure doses. The experimental results are compared with DFT calculations and molecular dynamics simulations allowing us to make conclusions about interactions and structures within the material - which influence the mobility of molecules and clustering, the anode material expansion, and the resulting performance of the system for energy storage and in particular rechargeable batteries.

\section{Experimental and theoretical setup}

In Figure \ref{scm:1}, we describe the methodologies employed to obtain and characterize samples based on our previous investigations~\cite{Candia2021}. The roughness of the substrate plays a determinant role in the thickness restriction of AlF$_{3}$ interlayer aggregates and this phenomenon occurs due to diffusion from the stepped edges into the interlaminar space of the host material. Figure~\ref{scm:1}(a) shows a schematic lateral representation of two different quality types of HOPG used in the experiments, specifically HOPG with a low and high density of staggered edges, identified as `HOPG A' and `HOPG B' respectively, Bruker grade I and II. Figure~\ref{scm:1}(b) illustrates the deposition, intercalation, and characterization process performed by STM under ultra-high vacuum (UHV) conditions. On the other hand, Figures~\ref{scm:1}(c) and (d) show the experimental data obtained by RBS (represented by squares), accompanied by simulated spectra generated from fitting models, represented by lines. In addition, a general descriptive scheme encapsulating the various components used in the data fitting is included. The interleaving process originates predominantly from these staggered edges. As a result, in the context of an ideal defect-free graphite (not depicted here), intercalation would lack feasibility. In a more practical scenario, an `A' type HOPG sample would result in a thin interlayer, while a `B' type would lead to a thicker one. For our experiments, we selected substrates ascribed to the `HOPG A' type resulting in a 24~nm penetration depth of AlF$_3$. In the supporting information, you can access a summary of the relevant experimental results obtained through previously published RBS studies. For a more comprehensive perspective refer to Candia et al. \cite{Candia2021}.

\subsection{STM-UHV}

Highly ordered pyrolytic graphite (HOPG from Bruker, UK, 12 mm $\times$ 12 mm $\times$ 1 mm) substrates were used for all the experiments. The clean HOPG surfaces were obtained by the tape cleavage in air and immediately loaded into the secondary reaction chamber of the STM-UHV, to be transferred after depositions to the main chamber of the system. AlF$_{3}$ molecules (CERAC INC., Milwaukee, Wisconsin, USA, 99.5\%) were thermally deposited in the HOPG surface normal direction, from a Knudsen cell charged with the anhydrous salt heated at 900 K, 200 mm apart from the substrate. The deposition was achieved under UHV conditions (during evaporations at a pressure in the high range of 10$^{-10}$ mbar) by a rate in between $6\times 10^{-3}$ and $2\times 10^{-2}$~ML~s$^{-1}$, and keeping the substrate at room temperature (RT). STM imaging was performed using a homemade Beetle scanning tunneling microscope in a UHV chamber with a base pressure in the low 10$^{-10}$ mbar range. All STM measurements were acquired at room temperature in the constant current mode using electrochemically etched tungsten (W) tips, with bias voltages and tunneling currents between 200 to 800 mV and 0.001 to 0.02 nA, respectively. These polycrystalline W tips were routinely cleaned by Ar$^{+}$ ion bombardment in UHV. Acquisition and image processing was performed using the WS$\times$M free software.\cite{Horcas2007}

\subsection{DFT}

\textit{Ab initio} calculations on the framework of DFT were performed using OpenMx3.9 package (Open source package for Material eXplorer), \cite{OPEN1,OPEN2} which incorporates norm-conserving pseudopotentials and pseudo-atomic localized orbitals (PAOs). The electronic exchange-correlation effects were treated within the generalized gradient approximation (GGA) as the functional proposed by Perdew, Burke, and Ernzerhof (PBE).\cite{PBE} We used the DFT-D3 approach for the correction of van der Waals interactions \cite{DFTD3}. Basis functions were created using a confinement scheme and labeled as: Al7.0-s2p2d2, C6.0-s2p2d1, and F6.0-s3p3d2f1 where Al, C, and F denote the chemical element, followed by the cutoff radius (Bohr radius), and the last set of symbols represent the primitive orbitals e.g. p2 indicates the use of two orbitals for the p component. 

 A cut-off energy of 300~Ry in the numerical integration and solution of the Poisson equation and a \textbf{k}-point mesh of 3$\times$4$\times$1 was used for the self-consistency calculation. To study the changes in the graphite surface under AlF$_3$ intercalation, we employed three graphite layers ---noted as L$_1$, L$_2$, and L$_3$ that adopt a Bernal AB stacking structure type for the HOPG--- to model six systems which we denote as, (i) AlF$_{3}|$$_{0}^{1}$, (ii) AlF$_{3}|$$_{0}^{2}$, (iii) AlF$_{3}|$$_{0}^{3}$, (iv) AlF$_{3}|$$_{1}^{2}$, (v) AlF$_{3}|$$_{2}^{2}$, and (vi) AlF$_{3}|$$_{1}^{3}$, where the superscript indicates the number of molecules intercalated between layers L$_{1}$ and L$_{2}$ and the subscript the number of molecules intercalated between layers L$_{2}$ and L$_{3}$ (see Figure \textbf{S2} in the Supporting Information). The orthorhombic supercells have dimensions of a= 1.704~nm (\textbf{x}-axis), b= 1.721~nm (\textbf{y}-axis), and c= 3.00 nm (\textbf{z}-axis), with 336 carbon atoms belonging graphite. The interlayer initial spacing was set as 0.4 nm. Full structural relaxation of atomic positions was performed up to a convergence force below 0.02 eV/\AA.  

The redistribution of charge density induced by the interaction between graphite and AlF$_{3}$ molecules was defined as follows $\Delta \rm D =D_{AlF_{3}|_{i}^{j}}-D_{AlF_{3}}-D_{Gr}$, where $\Delta$D is the difference charge density, D$_{\rm AlF_{3}|_{i}^{j}}$ is the charge density of each one of the aforementioned systems, D$_{\rm AlF_{3}}$ and D$_{\rm Gr}$ are the charge densities of the isolated AlF$_{3}$ and pristine graphite molecules, respectively. The charge transfer between graphite and AlF$_{3}$ molecules was calculated via a Mulliken population analysis. 

\subsection{MD}

In our atomistic model, a graphite region of 25~nm $\times$ 25~nm and seven layers of carbon atoms was filled with AlF$_3$ molecules. Periodic boundary conditions were established in the plane, leaving the system free on the axis orthogonal to the carbon layers. The bottom-most layer of carbon atoms is rigid and plays the role of the bulk, while all others are allowed to thermally move at 300~K. Thus bottom-most layer supports and mechanically stabilizes the other layers. The AlF$_3$ molecules are intercalated in the same number between the top six layers. There are no AlF$_3$ molecules present between the bottommost carbon layer and the subsequent mobile one. If there are no AlF$_3$ molecules situated in other carbon layers, it is due to the selection of the cut position. The density of AlF$_3$ molecules per layer varies between 0.015 and 0.03~AlF$_3$/nm$^2$. We opted for these densities to align with the overall apparent area of the blisters observed in the experiments, namely Figures~\ref{fgr:1_STM}(a) and (b).

The interatomic forces within graphite were derived using the appropriate Airebo potential~\cite{Steven2000}. Interactions between AlF$_3$ molecules were modeled using Born-Mayer potential parameters~\cite{Chaudhuri2004}. The adhesion forces between the carbon atoms in graphite and AlF$_3$ molecules are modeled as pure van der Waals interaction with parameters $\epsilon=6.68$~meV and $\sigma=3.166$~\r{A}, chosen in accordance with energies and separations obtained from DFT.

The molecular dynamics (MD) simulations were performed using time steps of 0.5~fs and an NVT thermostat in LAMMPS, a commonly used distributed classical MD code~\cite{LAMMPS}. The results are shown after 1~ns when the energy and structure of the system stopped evolving.

\section{Results and Discussion}
In GICs, it is known that invited species (guests) do not intercalate simultaneously in all the interlayer spaces of the host material. Instead, they do it according to a unique modulated pattern called {\it staging}.~\cite{Bhauriyal2017,Safran1987} The stage number $n$ is defined as the number of pristine layers of the material separating two consecutive intercalated interlayer spaces. Therefore, in {\it stage I} layers of graphite and guest molecules alternate one by one, in {\it stage II}, two pristine graphite layers separate two guest molecules layers, in {\it stage III}, three pristine graphite layers separating two guest molecules layers, and so on. In general, for any GICs, the arrangement of the guest in the layers is not perfect since, depending on the interaction between the host material and the guest, different configurations can be observed, such as the formation of small domains of guest material between the layers. In GICs, the most studied models to explain the intercalation mechanisms are those of R\'{u}dorff-Hoffman (RH)~\cite{ModelR} and Daumas-Herold (DH).~\cite{ModelDH} The RH model proposes a sequential filling in which empty and guest-baring layers are obtained alternatively without structural distortions of the graphite sheets. In contrast, the DH model proposes that the guest can be intercalated between all layers of the host by deforming the material. Some authors have proposed the coexistence of different stages at the same time, i.e. {\it staging stage} model.~\cite{MixModel} This could explain some anomalies observed in studies of lithium-ion intercalation in graphite from X-ray diffraction and entropy measurements for stages I and II experiments. \cite{YAZAMI2006312,SENYSHYN2015235}

So far, it has been demonstrated experimentally and theoretically that the intercalation of AlF$_3$ in graphene generates deformations or "blisters" on the surface. \cite{Candia2021} In this work, we will focus on the intercalation mechanism of AlF$_3$ in HOPG, and to facilitate the discussion of results, we will divide this section into two parts: (i) \textit{Morphology of AlF$_3$ blisters in HOPG}, where we present the experimental arrangement of the blisters on the graphite surface using STM, (ii) \textit{Revelation of the arrangement of blisters in HOPG}, where we make use of theoretical DFT calculations and MD simulations to discuss the modeled topography of three and seven layers of graphite with intercalated molecules. We compare the experimental and theoretical results to determine the number of AlF$_{3}$ molecules that form the blisters. At the end of this section, we propose a potential mechanism of the AlF$_{3}$ molecules intercalation.

\subsection{Morphology of the AlF$_3$ blisters in HOPG}

\begin{figure}[h!]
\centering
  \includegraphics[width=1\columnwidth]{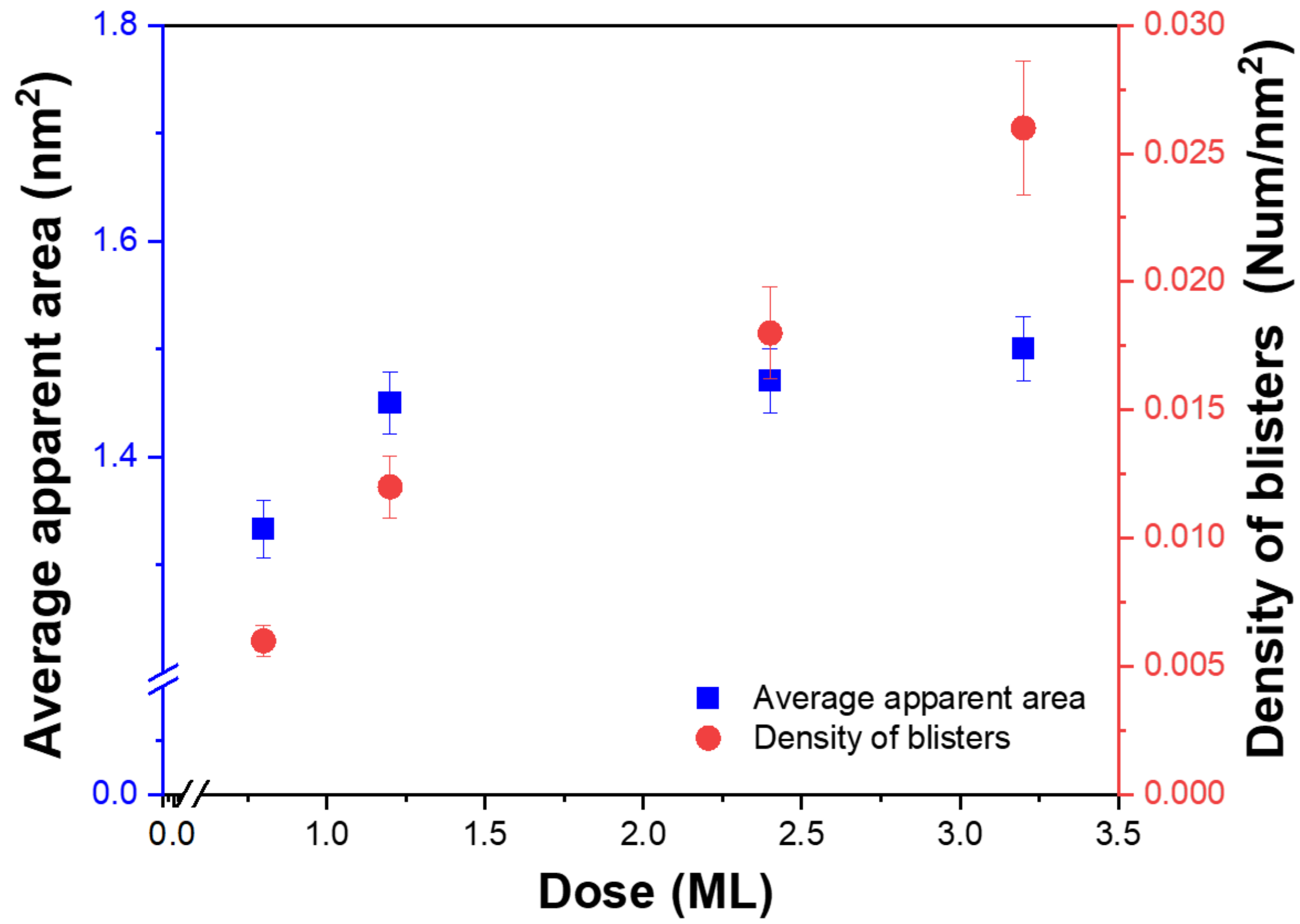}
  \caption{Variation of the blisters density and their average apparent area as a function of the AlF$_3$ molecules dose deposited on a HOPG substrate at 300 K.}
  \label{fgr:2_STM}
\end{figure}

In order to experimentally explore the topography and electronic characteristics of the intercalation of AlF$_3$ molecules between the HOPG layers, we performed STM images of the system at RT and different exposure doses. The deposition rate of AlF$_3$ molecules was varied between $6\times 10^{-3}$ and $2\times 10^{-2}$~ML~s$^{-1}$, while the coverage ranged from 0.8 to 3.5~ML. Our experimental analysis is based on data collected from four different experiments; however, in this section, we only present the corresponding result at those doses comparable with the results obtained with molecular dynamics. In Figures \ref{fgr:1_STM}(a) and \ref{fgr:1_STM}(b), we show the topography STM images of HOPG surfaces after the deposition of 1.2 and 3.0~ML of AlF$_3$, respectively. The results for the other two doses can be found in the Supporting Information. In both images, we observe bright regions on a dark brown background, corresponding to blisters generated by clusters of AlF$_{3}$ molecules intercalated with an inhomogeneous distribution beneath the topmost graphite layer. \cite{Candia2021} The density of these clusters is observed to be higher for the sample that has been more exposed, indicative of a correlation between intercalation density and exposure dose. Comparing both images, it is apparent that the size of the blisters varies with the deposited dose. In addition, it is observed that although the blister distribution across the graphite surface seems to be random for both doses, we found clusters of blisters characteristically located along a straight line, which we have indicated by the black dashed-line rectangles in the Figures \ref{fgr:1_STM}(a) and \ref{fgr:1_STM}(b). To clarify both, the size and the positions of the blisters, using the free software WS$\times$M, \cite{Horcas2007} a topographic flooding-type procedure is applied to images of Figures \ref{fgr:1_STM}(a) and \ref{fgr:1_STM}(b), which are shown in Figures \ref{fgr:1_STM}(c) and \ref{fgr:1_STM}(d), respectively. This procedure allows us to identify, count and measure the clusters of AlF$_3$ molecules, highlighting the areas that are above a threshold defined by the program. In our case, we set this threshold to 0.25 nm enough to minimize the noise inherent in image acquisition of the images and highlight only the AlF$_3$ clusters without affecting morphological characteristics. For ease of visualization, a color scheme was chosen to represent the cluster height. Thus, focusing preferentially on clusters within the dashed line rectangles, values between 2.5 - 3.0 nm and 0.3 - 0.5 nm are obtained for the diameters and apparent heights, respectively. At the same time, it is clearly observed how the average distance between the nearest blisters in each correlated set (we compared between dashed line rectangles in Figures \ref{fgr:1_STM}(a) and \ref{fgr:1_STM}(b)) decreases from 5 to 1.5 nm when the AlF$_3$ dose is increased. Figure \ref{fgr:2_STM} shows the evolution of the density of blisters (number per area, nm$^{-2}$) formed on the substrate and their average apparent area as a function of dose.  As expected, the blister density cells on the surface increase monotonically with the dose. The trend for the average apparent area of a blister is different. There is a significative increase at low doses (from 0.8 to 1.2 ML), followed by a plateau in the size of a single blister.

Given that STM solely focuses on the topography of the outermost atomic layer of the system, we are faced with a substantial challenge to achieve a definitive understanding of the intercalation dynamics. While our recent publications have confirmed that AlF$_{3}$ molecules integrate into the substrate via step edges and surface defects, inducing local separation between the carbon layers of HOPG without forming chemical bonds, yet influencing the local density of states (LDOS), we are still in a stage of partial understanding regarding the dynamics of this intercalation process. Therefore, in the following sections, we will address this topic in depth. Utilizing density functional theory (DFT) and molecular dynamics (MD) calculations, we will present a comprehensive interpretation of the intercalation mechanism that aligns with results obtained from scanning tunneling microscopy (STM) on the surface.

\begin{figure*}[ht!]
\centering
\includegraphics[width=2.0\columnwidth]{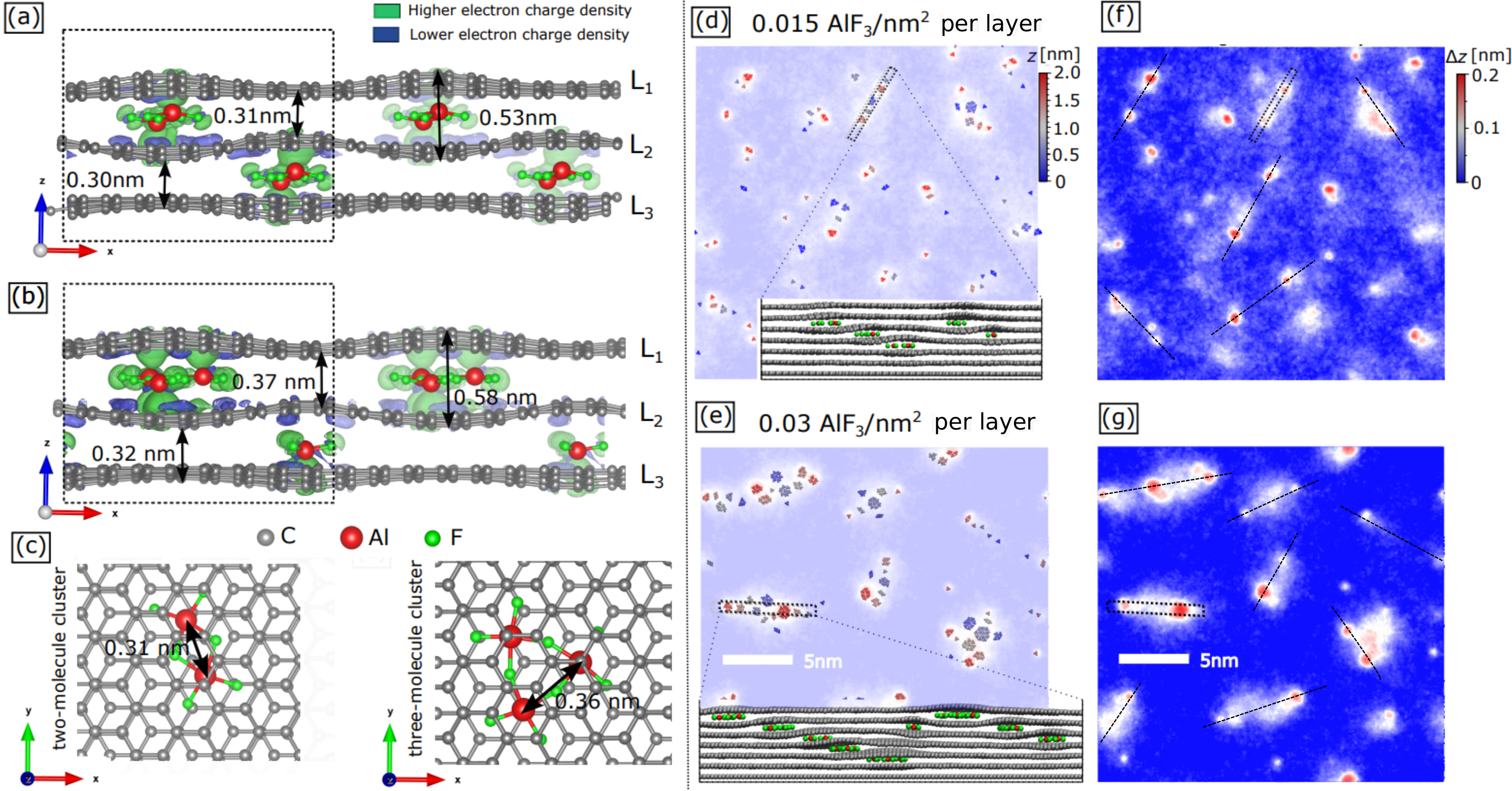}
\caption{(a)-(c) are the theoretical results from DFT. (a) and (b) are the geometric optimizations and difference charge density  for AlF$_{3}|$$_{2}^{2}$ and AlF$_{3}|$$_{1}^{3}$ systems, respectively, where the superscript indicates the number of molecules intercalated between layer L$_{1}$ and L$_{2}$ and the subscript the number of molecules intercalated between layer L$_{2}$ and L$_{3}$. Green regions correspond to D$>$0, i.e. higher electron charge density due to intercalation, and blue regions correspond to D$<$0, i.e. lower charge density due to intercalation. (c) top view of the AlF$_{3}|$$_{2}^{2}$ and AlF$_{3}|$$_{1}^{3}$ systems. (d)-(g) are the theoretical results from MD. (d) and (e) show the top and side view of the spatial arrangement of the clusters for two densities of AlF$_{3}$ molecules per layer of 0.015 AlF$_{3}$/nm$^2$ and 0.03 AlF$_{3}$/nm$^2$, respectively ---in order to facilitate visualization, we have established the color of the clusters, according to the height at which they are on the z-axis---. (f) and (g) show the results of the topography of the last layer after the formation of the clusters intercalated for the same density per layer. }
  \label{fgr:Teo_DFT_MD}
\end{figure*}

\subsection{Revealing the arrangement and size of the blisters in HOPG}

To understand the inherent dynamics of the mechanism of intercalation and to determine if any of the alternative RH, DR, or mixed-stage models fit the system under study, DFT calculations, and MD simulations are performed from modeling three and seven layers of graphite with intercalated AlF$_{3}$ molecules, respectively.  Figures \ref{fgr:Teo_DFT_MD}(a), \ref{fgr:Teo_DFT_MD}(b), and \ref{fgr:Teo_DFT_MD}(c) show the results obtained for the geometric optimization of two intercalated systems by DFT. In Figures \ref{fgr:Teo_DFT_MD}(a) and \ref{fgr:Teo_DFT_MD}(b), we depict side views for the AlF$_{3}|$$_{2}^{2}$ and AlF$_{3}|$$_{1}^{3}$ systems respectively, along with a top view in Figure \ref{fgr:Teo_DFT_MD}(c) showing the relaxed geometry of the clusters with two and three molecules involved in each of the aforementioned models. In addition, Figures \ref{fgr:Teo_DFT_MD}(a) and \ref{fgr:Teo_DFT_MD}(b) show superimposed the charge density differences induced by the interaction between the molecules and the graphite, represented by blue and green shaded regions. Figures \ref{fgr:Teo_DFT_MD}(d) and \ref{fgr:Teo_DFT_MD}(e) provide the MD results where the top and side views of the spatial arrangement of the clusters are observed for two densities of molecules per layer 0.015 and 0.03 AlF$_{3}$/nm$^2$, respectively. To facilitate visualization, we have established a color scale for the clusters, according to the height at which they are on the \textbf{z}-axis, set to zero at the bottom-most graphite layer. On the other hand, Figures \ref{fgr:Teo_DFT_MD}(f) and \ref{fgr:Teo_DFT_MD}(g) show images as a color map of the topography of the topmost layer produced by the intercalated clusters shown in Figures \ref{fgr:Teo_DFT_MD}(d) and \ref{fgr:Teo_DFT_MD}(e). In this case, the height difference ($\Delta$\textbf{z}) is measured from the level of the topmost unperturbed graphite layer. Since intercalation mechanisms are complex processes, we will discuss our results based on (1) \textit{In-plane interactions}, defined as those produced between molecules located in the same layer of the host material, and (2) \textit{Interlayer interactions}, established between molecules located in different layers of the host material.

\subsubsection{In-Plane interactions}

From DFT calculations, we determined that the intercalation process causes a charge transfer up to 2.63~e$^{-}$ from the graphite to the molecules for three-molecule clusters (see more details in the Supporting Information, section 3A). The electronic transfer redistributes the charges by polarizing the graphite sheets, which leads to the formation of local "transverse dipoles" between the molecules and the graphite, where molecules gain electrons while the graphite loses them. The transverse dipoles calculated by DFT, are visualized in the shaded regions of Figures \ref{fgr:Teo_DFT_MD}(a) and \ref{fgr:Teo_DFT_MD}(b), where the green regions correspond to the different charge density $\Delta$D$>$0, i.e., higher electron density after the charge transfer due to intercalation, and the blue regions correspond to difference charge density $\Delta$D$<$0, i.e., lower charge density upon intercalation \cite{Lin2015}\footnote{In the intercalation process, graphite gives electrons to the AlF$_{3}$ molecule, indicative of the electrochemical oxidation of graphitic carbons due to the intercalation of AlF$_{3}$, an average voltage of 3.18 V and 3.44 V for stage-2 and stage-1 was reported in Ref. [29]. These values stand out significantly compared to AlCl$_{4}$ anion-based batteries, which exhibit an operational voltage ranging between 2 and 2.5 V [15]. It is noteworthy that an increased operational voltage assumes paramount importance within rechargeable battery systems, owing to its intrinsic correlation with battery capacity, performance, and safety aspects. In the context of high-performance lithium batteries, an approximate operational voltage of 3.7 V is commonly observed. With regard to diffusion, our previously published theoretical study [29] highlighted the presence of low diffusion energy barriers associated with AlF$_{3}$ intercalants. This characteristic is poised to accelerate the charging process in a potential battery predicated upon this solvent}. These dipoles, which have components mainly on the \textbf{z}-axis, locally separate the graphite layers and elastically deform the material, playing a relevant role in the interactions between molecules in the same \textbf{xy}-plane, as described below (for more details on the theoretically calculated heights, see Table S1 of the Supporting Information).

According to DFT and MD theoretical results, the AlF$_{3}$ molecules approach each other forming small clusters parallel to the graphite \textbf{xy}-plane. Each intercalated molecule extends the C-C $\sigma$ bonds of graphene up to 2\%. A way for the system to reduce elastic energy is by pushing individual molecules into clusters. In turn, the elastic deformation of graphite results in an effective attractive force between molecules in the same plane. The pressure generated by the deformation of the layers is transferred into the internal pressure of the blisters.~\cite{Wang2016Nature,Yue2012} In this sense, the deformation of graphite drives the kinetics of the clustering and coalescence of AlF$_3$ molecules at a given graphite plane. Since the blister obtained theoretically and experimentally have a height and radii of less than 1.0 and 1.5~nm, we estimate the pressure inside the blisters from the linear plate model ~\footnote{Yue \textit {et al.} [47] used several models to analyze the adhesion energy between graphene layers: the membrane, the nonlinear plate, and the linear plate models. They report that the membrane model applies to large graphene blisters (height $>$1 nm), while the nonlinear plate model is generally more accurate and suitable for blisters of all sizes. The linear plate model is an approximation of the nonlinear plate model for small-sized blisters, which generally have heights below 0.3 nm and therefore are lower than their radii.}, using the expression of eq. S2 (see more details in the Supporting Information). According to this model, the pressure, expressed in terms of the adhesion energy, is inversely proportional to the square of the radius of the blister. \cite{Yue2012} Consequently, the pressure is higher in small blisters, and the pressure difference drives the molecules to diffuse from the smaller blisters to the larger ones. In this work, with the linear plate model and DFT, we estimate a pressure of 6.40, 4.06, and 3.01 GPa for blisters generated by intercalation of groups of one, two, and three AlF$_3$ molecules, respectively ---these values are on the order of the pressures reported by Wang \textit{et al.} \cite{Wang2016Nature} and Villareal \textit{et al.} \cite{Villarreal2021} for blisters with radii smaller than 1 nm---. It is important to note both from experimental (Figure \ref{fgr:2_STM}) and theoretical results (Figures \ref{fgr:Teo_DFT_MD}(f) and \ref{fgr:Teo_DFT_MD}(g)),  that no \textbf{xy}-plane contains blisters with diameters larger than 3 nm. This appears to indicate the existence of a critical coalescence value. This outcome is mainly due to elastic deformation and the generation of transverse dipoles by the molecules between adjacent layers of the graphite. This critical coalescence number could explain why the average area of the blisters remains almost constant for doses higher than 1.2 ML, as shown in Figure \ref{fgr:2_STM}. 

By analyzing the STM images and our DFT and MD results, we have estimated that blisters with diameters in the range of 1.3 to 2.0 nm are made up of clusters of two to six AlF$_{3}$ molecules, respectively. The size of the blisters formed between the L$_{1}$ and L$_{2}$ layers depends on the concentration of intercalated molecules. According to the results obtained by MD, when the density of molecules per layer is 0.015 AlF$_{3}$/nm$^2$, the blisters on the surface have smaller diameters. In contrast, for a density of 0.03 AlF$_{3}$/nm$^2$, the diameters increase. This is consistent with the experimental data presented in Figure \ref{fgr:2_STM}, which indicates that the most significant increase in the average area of the blisters occurs for doses lower than 1.2 ML.

Regarding the height of the blisters, and according to the topography modeled by calculation with three graphite layers by DFT and seven layers by MD, the clusters deform the graphite layer such that the blister reaches the apparent height of 0.22 nm (see Figure \ref{fgr:Teo_DFT_MD}). It should be noted that experimentally the apparent height could increase, since, on the one hand, the innermost clusters could push everything upwards and, on the other hand, there could be an overestimation during acquisition due to unknown tip conditions.\cite{Wang2016Nature}

\subsubsection{Interlayer interactions}

Figures~\ref{fgr:Teo_DFT_MD}(d) and \ref{fgr:Teo_DFT_MD}(e) depict the spatial arrangement of clusters along the \textbf{z}-axis. The atoms of the molecules are colored according to their vertical position, from blue for the molecules in the lower layer ($z=0$~nm) to red for those in the upper layer ($z=2$~nm). According to the results, it is not observed for any of the two densities of AlF$_{3}$ molecules that the average size of the clusters varies with the depth. This observation is evident in Figure~\ref{fgr:Teo_DFT_MD}(d), where the clusters in the lower layers (represented in blue) exhibit similar sizes to those in the upper layers (represented in red).
However, the most striking feature is the alignment of the clusters obtained in the MD simulations. The topography images of the MD simulations in Figures \ref{fgr:Teo_DFT_MD}(f) and \ref{fgr:Teo_DFT_MD}(g), clearly show the formation of blisters of different diameters in the topmost layer, which are aligned and separated about 4 nm apart from each other in the \textbf{xy}-plane. The self-organization of clusters in a line is observed for both modeled densities. The alignment direction of the blisters is random and does not correlate to any crystallographic orientation of the HOPG, as indicated by the dashed boxes in the molecular dynamics figures. Such random alignment is also evident in the experimental STM image, see the dashed line box in Figure \ref{fgr:1_STM}(b). A relevant aspect to highlight from these MD results is the role played by the innermost clusters since how the blisters are aligned on the surface depends on this, see the inset of Figures \ref{fgr:Teo_DFT_MD}(d) and \ref{fgr:Teo_DFT_MD}(e). Below the surface, the formation of clusters between the different layers prevents larger structures from forming within a given layer. As a result of this interlayer self-assembly blisters in the topmost layer remain approximately 4 nm apart from each other, despite the elastic forces that attempt to coalesce them, see Figure \ref{fgr:Teo_DFT_MD}(f). 
Since the results of the MD simulations were able to reproduce the experimental characteristics of the blisters in terms of their alignment and size distribution, they can be employed to gain a deeper understanding of the processes taking place within the clusters below the surface. According to MD, the clusters between neighboring layers do not stack vertically. The reason behind this is the elastic deformations of graphite and the repulsive interaction of local transverse dipoles. The insets in Figures \ref{fgr:Teo_DFT_MD}(d) and \ref{fgr:Teo_DFT_MD}(e), which show a side view of the highlighted regions, reveal the depth distribution of the clusters that form blisters observed in the topmost layer (shown in Figures \ref{fgr:Teo_DFT_MD}(f) and \ref{fgr:Teo_DFT_MD}(g), respectively). Each region framed by the dashed line boxes represents a superstructure formed by intercalated clusters, and displaced laterally from each other, between different pairs of layers. A "local staging" is observed in each superstructure, indicating the coexistence of "mixed stages". For instance, for the dilute system of 0.015 AlF$_{3}$/nm$^2$, the organization of the molecules combines stages III and IV, while for the densest system of 0.03 AlF$_{3}$/nm$^2$, it combines stages IV and V. This arrangement gives rise to an aligned distribution of the clusters and explains the average distance of 4 nm between blisters at the topmost layer in Figures \ref{fgr:Teo_DFT_MD}(f) and Figures \ref{fgr:Teo_DFT_MD}(g), and also in the STM image of \ref{fgr:1_STM}(b) corresponding to the experiment.

\section{Conclusions}

In this work, we employed a combination of scanning tunneling microscopy (STM), density functional theory (DFT) calculations, and molecular dynamics (MD) simulations to gain insight into the process of AlF$_{3}$ intercalation in highly oriented pyrolytic graphite (HOPG). Experimentally, we investigated the intercalation mechanism of thermally dosed AlF$_{3}$ molecules perpendicularly to the HOPG surface at room temperature under ultra-high vacuum conditions (in the high range of 10$^{-10}$~mbar). STM images obtained for varying molecule doses revealed that the blisters are not uniformly distributed over the graphite surface. Besides, it was observed that some blisters align locally with random orientations and without following any preferential crystallographic direction, see the line and dashed rectangle in Figure \ref{fgr:1_STM}(b). Since STM only probes the surface of the system, we supplemented our experimental results with theoretical DFT calculations and MD simulations, which allowed us to propose a plausible explanation for the local alignment of the blisters observed on the surface. 

We used DFT and MD simulations to study in-plane and interlayer interactions between AlF$_{3}$ molecules and graphite. The in-plane interactions involve the arrangement of AlF$_3$ molecules between two graphite layers. The findings include the formation of AlF$_3$ molecule clusters resulting in localized elastic deformation graphite layers. Charge transfer between graphite and AlF$_3$ molecules induces the formation of transverse dipoles. The molecules tend to form clusters between two graphite layers, driven by pressure differences. The pressures stemming from adhesive van der Waals forces between layers of 6.40, 4.06, and 3.01 GPa are estimated from DFT calculations for blisters containing one, two, and three intercalated molecules, respectively. The interlayer ordering of clusters reveals that clusters formed in deeper layers of the material direct the alignment of surface blisters. These three-dimensional superstructures, extend several layers into the depth of the material and connect through the bulk of material surface blisters arranged in a local line. The movement of observed surface clusters is therefore constrained by neighboring deeper clusters. Our model-based findings provide insights into the arrangement and behavior of blisters on the material's surface.

An important finding is the existence of superstructures visible in the experiment as surface blisters arranged in a local line and separated by an average of 4 nm. Every cluster affects the layers around it. Due to neighboring clusters in graphite layers under or above, AlF$_3$ clusters are quenched together and cannot get closer or grow. They cannot also stack on top of each other since the stacking would result in a sharp increase in elastic energy required to deform graphite layers. Therefore AlF$_3$ clusters avoid stacking on top of each other. As a result, the clusters group laterally next to each other - sharing the deformation of graphene to reduce elastic energy. Daumas and Herold postulated a model, in which the graphene layers are flexible and deform around clusters of the intercalated species.\cite{ModelDH} In the R\'{u}dolf-Hoffman model, layers are not elastic, and molecules intercalate by separating these layers. \cite{ModelR} Our results underline the effect of the elasticity (deformability) of the host material (graphite) on the evolution of intercalation.  Based on our results, clusters arrange via elastic interactions in graphene. The local deformations of graphene sheets and resulting local strains lead to mixed stages since neighboring clusters are separated by one layer of graphene, while simultaneously superstructures are separated by several nanometers. We can deduct what is an outcome of this process at higher densities: one expects the growth of clusters while they avoid vertical stacking. The process will continue until distances between the AlF$_3$ clusters sufficiently decrease and the density of AlF$_3$ within one layer becomes sufficiently high. At this point, the clusters will coalesce into the full AlF$_3$ layers removing elastic strain on host material (graphite) layers.

\section{Supporting Information}
 The supporting information is divided into three sections, the first is a summary of previous results by RBS of the system under study. The second section shows the experimental STM images of the HOPG before and after evaporation of 0.8 and 3.2 ML of AlF$_3$. Finally, in the third section, we present the computational details of the six models studied by DFT, reporting the formation energies, charge transfer, and some distances of interest between the molecules. In addition, we include a detailed calculation of the pressures of the blisters formed by one, two, and three intercalated molecules.
 
\section{Acknowledgement}

The authors acknowledge the financial support by the Consejo Nacional de Investigaciones Cient\'ificas y T\'ecnicas (CONICET) through grant PIP 2021-0384 and the Agencia Nacional de Promoci\'on Cient\'ifica y Tecnol\'ogica (ANPCyT) through the grant PICT-2019-04545 and Universidad Nacional del Litoral through grant CAI+D 2020-50620190100016Li. The present work used computational resources of the Piray\'u cluster, funded by the Agencia Santafesina de Ciencia, Tecnolog\'ia e Innovaci\'on (ASACTEI), Province of Santa Fe, Argentina, through grant AC-00010- 18, resolution N$^{\circ}$117/14. This equipment is part of the National System of High-Performance Computing of the Ministry of Science and Technology, Argentina. I.S. acknowledges the support of the Ministry of Education, Science, and Technological Development of the Republic of Serbia through the Institute of Physics Belgrade. Molecular dynamics calculations were run on the PARADOX supercomputing facility at the Scientific Computing Laboratory, Center for the Study of Complex Systems of the Institute of Physics Belgrade. Last but not least, the authors acknowledge the financial support of the European Commission through the project ULTIMATE-I, grant ID 101007825. 

\bibliography{RI-arxiv}

\end{document}